\documentclass[12pt]{extarticle} 
\usepackage[cp1251]{inputenc}
\usepackage{graphicx}
\usepackage{amssymb,amsmath}
\usepackage{mathrsfs}
\usepackage{textcomp}
\usepackage{cite}
\usepackage{color}
\usepackage{ulem}
\usepackage{setspace} 
\def\@oddhead{\thedate}

\newcommand{\p}{\mspace{1mu}}\newcommand{\pp}{\mspace{2mu}}
\newcommand{\ppp}{\mspace{3mu}}\newcommand{\pppp}{\mspace{4mu}}
\newcommand{\mm}{\mspace{-2mu}}
\newcommand{\mmm}{\mspace{-3mu}}

\definecolor{dblue}{rgb}{0.,0.,0.7}\definecolor{dred}{rgb}{0.4,0.,0.}

\newcommand{\lt}{\left}\newcommand{\rt}{\right}

\newcounter{seksjon}

\title{\bf{\Large
ON THE THEORY OF A NON-LINEAR NEUTRAL SCALAR FIELD WITH SPONTANEOUSLY BROKEN SYMMETRY %
\footnote{Published in The Journal of Kharkov National University
\textbf{859}, Physical series ``Nuclei, Particles, Fields'',
Iss.~2/42/, p. 9-20 (2009).
}}
}

\author{Yu.M. Poluektov
\\
\small National Science Center \\
\small ``Kharkov Institute of Physics and Technology''\\
\small 1, Akademicheskaya St., 61108 Kharkov, Ukraine\\
\small E-mail: yuripoluektov@kipt.kharkov.ua
\date{} 
}

\begin{document}
\maketitle

\renewcommand\abstractname{}
\begin{abstract}
On the example of a real scalar field, an approach to quantization
of non-linear fields and construction of the perturbation theory
with account of spontaneous symmetry breaking is proposed. The
method is based on using as the main approximation of the
relativistic self-consistent field model, in which the influence of
vacuum field fluctuations is taken into account when constructing
the one-particle states. The solutions of the self-consistent
equations determine possible states, which also include the states
with broken symmetries. Different states of the field are matched to
particles, whose masses are determined by both parameters of the
Lagrangian and vacuum fluctuations. The density of the vacuum energy
in these states is calculated. It is shown that the concept of
Bogolubov's quasi-averages can naturally be applied for definition
of exact Green functions in the states with broken symmetries.
Equations for exact one- and two-point  Green functions are obtained.
\end{abstract}

\newpage

\noindent \textbf{INTRODUCTION} %

The study of states with broken symmetries in quantum field theory
was initiated under influence of the works on microscopic theory of
superfluidity \cite{1} and superconductivity \cite{2,3}, in the
works of Nambu and Jona-Lasinio\cite{4}, Wax and Larkin \cite{5},
Goldstone\cite{6} et al. The effect of spontaneous symmetry breaking
is an important element of contemporary theory of elementary
particles \cite{7,8,9}. Most often it is performed by way of
including of massive scalar fields with an ``incorrect'' sign of the
square of mass into the Lagrange function (Goldstone mechanism
\cite{6}). This mechanism was used in construction of a unified
theory of weak and electromagnetic interactions \cite{7}. Another
way to describe spontaneous symmetry breaking was suggested by Nambu
and Jona-Lasinio \cite{4}, Wax and Larkin \cite{5} on the example of
chiral symmetry breaking in hadron physics. The method used in the
works \cite{4,5} is based upon the analogy with the
superconductivity theory, and it got the name of dynamical mechanism
of symmetry breaking \cite{10}. Both the Goldstone mechanism and the
dynamical one are the ways of description of the same phenomenon.

The aim of the present work is to construct on the example of a real
scalar field a common approach for description of the states of
non-linear relativistic quantized fields with arbitrary
spontaneously broken symmetries. The proposed approach to the
description of spontaneous symmetry breaking in the quantum field
theory generalizes to the relativistic case the quantum-field
approach for microscopic description of non-relativistic
many-particle Fermi- and Bose-systems with broken symmetries, that
is developed in the works \cite{11,12,13,14,15}, and is, in its
essence, closer to the dynamical mechanism.

In the developed approach the Lagrangian of a non-linear field is
split, as usually, into two parts, one of which describes free
particles and the other one their interaction. The peculiarity of
the proposed method consists in the way of construction of the
Lagrangian of non-interacting particles. This peculiarity consists
in choice of Lagrangian that contains the terms not higher than
quadratic in field operators, including those terms whose symmetry
is lower than the symmetry of the initial Lagrangian. The parameters
that enter into the free Lagrangian are selected from the
requirement of its maximal proximity to the full Lagrange function.
The system of non-linear equations derived based upon this
condition, which determines the parameters of the free Lagrangian,
in particular the particle masses, can have different solutions that
describe the states with different symmetries. Such an approach
lets, to a certain extent, account for interaction already in the
stage of construction of one-particle states. In particular, the
value of the interaction constant determines the particle masses
that turn out to be different in different states. The perturbation
theory constructed on the basis of the mentioned choice of the main
approximation can be also applicable in the case when the
interaction constant is not small. It is notable that, when using
the postulated method of construction of one-particle states, the
Hamiltonian of interaction automatically takes the form of normally
ordered product of field operators. Owing to this, many diagrams are
excluded from the perturbation theory expansion, namely all diagrams
containing loops, \emph{i.e.} the lines both ends of which are
connected to the same vertex and, therefore, the structure of the
perturbation theory is significantly simplified. In the present work
the equations for one- and two-point Green functions are also
obtained. It is shown that, when determining exact Green functions
in the states with spontaneously broken symmetries, it is natural to
use the concept of quasi-averages introduced by Bogoljubov \cite{16}
in statistical mechanics. Although in this work we used as an
example the real scalar field with broken discrete symmetry, the
proposed method of description of the states with broken symmetries
is applicable also for other, including non-linear fermionic, fields
\cite{23}.
\\ \\
\noindent \textbf{RELATIVISTIC MODEL OF SELF-CONSISTENT FIELD}

Let us consider a Bose system described by the secondary-quantized
Hermitian field operator $\varphi \equiv \varphi \lt(x\rt)$ that
depends on space coordinates $\mathbf{x}$ and time $t$, so that
$x\equiv {\kern 1pt} \, \, x_{\mu } =\lt({\bf x},it\rt)$, $\mu
=1,2,3,4$. The system is characterized by the density of Lagrange
function
\begin{equation} \label{GrindEQ__1_}
L=-\frac{1}{2} \lt[\lt(\partial_{\mu } \varphi \rt)^2 +\kappa ^2 \varphi ^2 \rt]-\frac{g}{4!} \varphi ^4 , %
\end{equation}
where $\partial_{\mu } =\partial/\partial x_\mu$ and the metrics
used is $ab=\textbf{ab}+a_4 b_4 =\textbf{ab}-a_{0} b_{0}$. It is
supposed that the sign of the real parameter $\kappa^2$ (that we
will call the mass parameter) is arbitrary, and the interaction
constant $g$ is positive. The system of units used in this paper is
$\hbar=c=1$. Let us note that, when constructing the quantum field
theory, usually in the initial step one considers classical fields,
and in the following step one realizes the ``quantization'' of the
classical fields by way of transition to the secondary-quantized
operators that obey certain commutation relations. Such an approach,
no doubt,  is justified in both historical and methodological
respects. However, in virtue of the fact that the quantum
description is a more profound level of description of the reality,
we will start immediately from it. Classical fields in this case
must be a limiting case of quantum fields.

For field operators at the same time arguments the standard
commutation relations are postulated:
\begin{equation} \label{GrindEQ__2_}
\begin{array}{c}
\lt[\pi \lt(\textbf{x},t\rt),\varphi \lt(\textbf{x}',t\rt)\rt]=-i\delta \lt(\textbf{x}-\textbf{x}'\rt),
\\
\lt[\varphi \lt(\textbf{x},t\rt),\varphi \lt(\textbf{x}',t\rt)\rt]=\lt[\pi \lt(\textbf{x},t\rt),\pi \lt(\textbf{x}',t\rt)\rt]=0,
\end{array}
\end{equation}
where $\pi\lt(\textbf{x},t\rt)=\partial
L/\partial\dot{\varphi}\mm\lt(\textbf{x},t\rt)$ is the canonically
conjugated momentum. The field obeys the non-linear operator
equation
\begin{equation} \label{GrindEQ__3_}
\partial_\mu\partial_\mu\varphi-\kappa^2\varphi-\frac{g}{3!}\varphi^3=0.
\end{equation}
The field operators in \eqref{GrindEQ__1_} are taken in the
Heisenberg picture:
\begin{equation} \label{GrindEQ__4_}
\varphi\lt(\textbf{x},t\rt)=e^{iHt}\varphi\lt(\textbf{x},0\rt)e^{-iHt},\quad\quad\pi\lt(\textbf{x},t\rt)=
e^{iHt}\pi\lt(\textbf{x},0\rt)e^{-iHt},
\end{equation}
where $\varphi\lt(\textbf{x},0\rt)$, $\pi\lt(\textbf{x},0\rt)$ are
operators in the Schr\"odinger picture and the Hamilton operator has
the form
\begin{equation} \label{GrindEQ__5_}
H=\int d\textbf{x} \lt\{\frac{1}{2} \lt[\pi^2 +\lt(\nabla\varphi\rt)^2+\kappa^2
\varphi^2\rt]+\frac{g}{4!}\varphi^4\rt\}.
\end{equation}

The transition from the field operators to the particle operators
for free fields is performed in a well-known way. For this purpose,
the field operator is split into the positive-frequency and
negative-frequency parts and transition to the Fourier
representation is carried out. The coefficients of such expansion
have the sense of creation and annihilation operators of
non-interacting particles. In the case of non-linear theory the
described prescription of introducing the one-particle states cannot
be used. In this case the coefficients of Fourier transform do not
have the sense of operators of creation and annihilation of states
with a correct relativistic law of dispersion. In the initial
Lagrangian one could select the part quadratic in the field
operators and consider it, as is usually done, as the free particles
Lagrangian; and consider the rest part of the Lagrangian as a
perturbation. It is worth noting that this ``obvious'' way of
constructing the one-particle states is, in its essence, a tacit
agreement. However, such a partition is not uniquely possible.
Moreover, it is not effective in the case of states with
spontaneously broken symmetries, for whose description the account
of the effects caused by nonlinearity of the system is essential.
Particularly, such consideration does not have sense when the mass
parameter $\kappa^2$ is negative, because in this case the free
Lagrangian describes the objects with ``incorrect'' (tachyonic) law
of dispersion. The possibility of introducing of tachyons in physics
and of giving them a real status was widely discussed earlier (see
the book \cite{17}). A need of introducing of such objects as
tachyons is under question, since, if a field is described by
non-linear equations, the presence of an ``incorrect'' sign of the
square of mass in the quadratic part of Lagrangian does not at all
indicate the existence of particles with exotic properties,
although, as was shown by Goldstone \cite{6}, this case describes
quite a real physical situation. When constructing a non-linear
field theory it is worth remembering, that the initial physical
sense is only inherent in the full Lagrangian of the system. The
splitting of the full Lagrangian into the sum of the Lagrangian of
``non-interacting'' particles and the interaction Lagrangian is,
obviously, ambiguous. Indeed, besides some splitting $L=L_1+L_2$,
one can proceed from another splitting $L=L'_1+L'_2$, where
$L'_1=L_1+\Delta L,\, L'_2=L_2-\Delta L$ and $\Delta L$ is some
operator addition. By fixing the way, in which we construct the part
of Lagrangian describing the one-particle states, we in fact give a
definition for the notion of ``non-interacting particle'' in the
framework of the non-linear theory. Now let us present the full
Lagrangian \eqref{GrindEQ__1_} as the sum
\begin{equation} \label{GrindEQ__6_}
L=L_{0}+L_C,
\end{equation}
where $L_{0}$ is the Lagrangian that contains the terms not higher
than quadratic in the field operators, including those which violate
the symmetry of the initial Lagrangian $L$. Practically the only
requirement imposed onto the form of the Lagrangian $L_{0}$, besides
its quadraticness, is its Lorentz-invariance. This Lagrangian, by
definition, will be considered as the Lagrangian of
``non-interacting'' particles. The second term, $L_C$, is the
Lagrangian that contains the terms not present in $L_{0}$. When
constructing the perturbation theory we consider the first term in
\eqref{GrindEQ__6_} as a non-perturbed Lagrangian, and the second
one -- as a perturbation describing the interaction of particles
determined by the first term. In the considered case of a real
scalar field
\begin{equation} \label{GrindEQ__7_}
L_{0} =-\frac{1}{2} \lt(\partial_{\mu } \varphi \rt)^2 +D\varphi ^2 +p\varphi -\Omega ,
\end{equation}
where $D,p,\Omega$ are real parameters. The interaction Lagrangian,
obviously, has the form
\begin{equation} \label{GrindEQ__8_}
L_C =L-L_{0} =-\frac{g}{4!} \varphi ^4 -\lt(\frac{\kappa ^2 }{2} +D\rt)\varphi ^2 -p\varphi +\Omega .
\end{equation}
In the main Lagrangian we have added and subtracted the operator
term $\Delta L=D\varphi ^2 +p\varphi -\Omega$ without changing $L$.
The parameters $D,p,\Omega$\, can be to a large degree arbitrary,
and one should choose them using some extra considerations. Let us
note that $L_{0}$ contains a term linear in $\varphi$, that breaks
the symmetry relative to the transformation $\varphi \to -\varphi$\,
inherent in the Lagrangian $L$. An analogical statement is also true
for $L_C$. The selected split into the main part and the interaction
takes place also for the Hamiltonian $H=H_{0}+H_C$:
\begin{equation} \label{GrindEQ__9_}
H_{0} =\int d\textbf{x}\lt[\frac{\pi ^2 }{2} +\frac{1}{2} \lt(\nabla \varphi \rt)^2 -D\varphi ^2 -p\varphi +\Omega \rt] ,
\end{equation}
\begin{equation} \label{GrindEQ__10_}
H_C =\int d\textbf{x}\lt[\frac{g}{4!} \varphi ^4 +\lt(D+\frac{\kappa ^2 }{2} \rt)\varphi ^2 +p\varphi -\Omega \rt] .
\end{equation}
As far as the full Hamiltonian is expressed in the same way both in
terms of Heisenberg and of Schr\"odinger operators, then $H_{0} $
and $H_C$ in \eqref{GrindEQ__9_} and \eqref{GrindEQ__10_} can also
be expressed in terms of the field operators in these pictures. Let
us consider more in detail the main approximation given by the
Hamiltonian \eqref{GrindEQ__9_}, expressed in terms of Schr\"odinger
operators. This approximation describes the system of
``non-interacting'', in our understanding, particles, in the
framework of the self-consistent field model. Let us define the
operators in the interaction picture:
\begin{equation} \label{GrindEQ__11_}
\hat\varphi\lt(\textbf{x},t\rt)=e^{iH_{0} t} \varphi\lt(\textbf{x},0\rt)e^{-iH_{0} t} ,\quad \hat\pi \lt(\textbf{x},t\rt)=e^{iH_{0} t} \pi\lt(\textbf{x},0\rt)e^{-iH_{0} t} .
\end{equation}
Simultaneous commutation relations for operators
\eqref{GrindEQ__11_} have the same form \eqref{GrindEQ__2_}, as for
Heisenberg operators. Obviously, $H_{0}$ is expressed in terms of
the operators \eqref{GrindEQ__11_} in the same way as in terms of
Schr\"odinger operators:
\begin{equation} \label{GrindEQ__12_}
H_{0} =\int d\textbf{x}\lt[\frac{\hat\pi^2 }{2} +\frac{1}{2} \lt(\nabla \hat\varphi\rt)^2 -D\hat\varphi^2 -p\hat\varphi+\Omega \rt] .
\end{equation}
Before reducing the Hamiltonian \eqref{GrindEQ__12_} to the diagonal
form it is necessary to eliminate the linear term from it by
carrying out the substitution
\begin{equation} \label{GrindEQ__13_}
\hat\varphi\lt(\textbf{x},t\rt)=\hat\psi\lt(\textbf{x},t\rt)+\chi ,
\end{equation}
where $\chi$ is a real parameter, $\hat\psi\lt(\textbf{x},t\rt)$ an
operator obeying the commutation rules analogous to
\eqref{GrindEQ__2_}. The need for a ``shift'' of the field operator
by the value $\chi$ is, apparently, conditioned by breaking of the
symmetry relative to the operation $\varphi \to -\varphi $. So,
following the terminology used in the theory of phase transitions we
will call this quantity the order parameter, and the state in which
$\chi \ne 0$ -- the ordered state. Following the analogy with the
non-relativistic theory of Bose-systems \cite{13}, the parameter
$\chi$ can also be called the condensate wave function, and the
operator $\hat\psi\lt(\textbf{x},t\rt)$ -- the over-condensate field
operator. The fields $\hat\varphi\lt(\textbf{x},t\rt)$ and
$\hat\psi\lt(\textbf{x},t\rt)$ are connected by the unitary
transformation $\hat\varphi\lt(\textbf{x},t\rt)=U_{\chi }^{+}
\hat\psi\lt(\textbf{x},t\rt)U_{\chi }$, where $U_{\chi } =\exp
\lt\{-i\chi \int d\p\textbf{x}'\p\hat\pi\mm\lt(\textbf{x}'\rt)\rt\}$
is a unitary operator.

Let us substitute \eqref{GrindEQ__13_} into \eqref{GrindEQ__12_} and
require that the terms linear in the operator
$\hat\psi\lt(\textbf{x},t\rt)$ drop from the obtained expression.
This leads to the condition
\begin{equation} \label{GrindEQ__14_}
p+2\chi D=0.
\end{equation}
As a result the Hamiltonian \eqref{GrindEQ__12_} takes the form
\begin{equation} \label{GrindEQ__15_}
H_0=\int d{\bf x}\lt[\frac{\hat\pi^2}{2} +\frac{1}{2}\lt(\nabla\hat\psi\rt)^2-D\hat\psi^2+D\chi^2+\Omega\rt].
\end{equation}
Let ${\lt| 0 \rt\rangle} $ be a vacuum vector of the system with the
Hamiltonian \eqref{GrindEQ__15_}. As far as $H_0$ does not contain
terms linear in $\hat\psi\lt(x\rt)$ and the Fock space is used, then
obviously ${\lt\langle 0 \rt|} \hat\psi{\lt| 0 \rt\rangle} =0$.
Therefore, by virtue of \eqref{GrindEQ__13_}:
\begin{equation} \label{GrindEQ__16_}
\chi ={\lt\langle 0 \rt|} \hat\varphi{\lt| 0 \rt\rangle} .
\end{equation}
Thus, the state with broken symmetry is characterized by a nonzero
vacuum average of the field operator.

Until now the parameters $D,p,\Omega$ were not fixed in no wise. One
should give a way to determine these parameters. By indicating it,
we will in fact determine the mode of construction of the
one-particle states in the framework of the developed approach. We
postulate that the parameters $D,p,\Omega$ are found from the
requirement that the approximating Hamiltonian $H_0$ be in some
respect maximally close to the exact Hamiltonian
\eqref{GrindEQ__4_}. In order to formulate this requirement
quantitatively, let us define the functional
\begin{equation} \label{GrindEQ__17_}
J\equiv {\lt\langle 0 \rt|} H-H_0 {\lt| 0 \rt\rangle}^2 ={\lt\langle 0 \rt|} H_C {\lt| 0 \rt\rangle}^2
\end{equation}
that characterizes the difference between the exact and
approximating Hamiltonians. The Hamiltonian $H_C$ in
\eqref{GrindEQ__17_} is taken in the interaction picture. The
average over the vacuum state of $H_C$ has the form
\begin{equation} \label{GrindEQ__18_}
{\lt\langle 0 \rt|} H_C {\lt| 0 \rt\rangle} =\int d\textbf{x}\lt[\frac{3g}{4!} \lt({\lt\langle 0 \rt|} \hat\varphi^2 {\lt| 0 \rt\rangle} \rt)^2 -\frac{2g}{4!} \chi ^4 +\lt(D+\frac{\kappa ^2 }{2} \rt){\lt\langle 0 \rt|} \hat\varphi^2 {\lt| 0 \rt\rangle} +p\chi -\Omega \rt].
\end{equation}
Let us require that the following conditions hold
\begin{equation} \label{GrindEQ__19_}
\frac{\partial J}{\partial {\lt\langle 0 \rt|} \hat\varphi^2 {\lt| 0 \rt\rangle} } =0,\quad \quad \frac{\partial J}{\partial \chi } =0,\quad \quad \frac{\partial J}{\partial \Omega } =0.
\end{equation}
Here one should consider the parameters $D,p,\Omega$\, being
independent and, for the present, not take into account the relation
\eqref{GrindEQ__14_}. From the conditions \eqref{GrindEQ__19_} we
obtain
\begin{equation} \label{GrindEQ__20_}
6g'{\lt\langle 0 \rt|} \hat\varphi^2 {\lt| 0 \rt\rangle} +D+\frac{\kappa ^2}{2} =0,
\end{equation}
\begin{equation} \label{GrindEQ__21_}
p=8g'\chi ^3,
\end{equation}
\begin{equation} \label{GrindEQ__22_}
\Omega =g'{\lt\langle 0 \rt|} \hat\varphi^4 {\lt| 0 \rt\rangle} +\lt(D+\frac{\kappa ^2 }{2} \rt){\lt\langle 0 \rt|} \hat\varphi^2 {\lt| 0 \rt\rangle} +p{\lt\langle 0 \rt|} \hat\varphi{\lt| 0 \rt\rangle} ,
\end{equation}
where, for the brevity, we set $g'\equiv g/4!$. Taking into account
the relation \eqref{GrindEQ__16_}, we find
\begin{equation*}
\begin{array}{c}
{\lt\langle 0 \rt|} \hat\varphi^2 {\lt| 0 \rt\rangle} ={\lt\langle 0 \rt|} \hat\psi^2 {\lt| 0 \rt\rangle} +\chi ^2, \qquad %
{\lt\langle 0 \rt|} \hat\varphi^4 {\lt| 0 \rt\rangle} =3{\lt\langle 0 \rt|} \hat\psi^2 {\lt| 0 \rt\rangle}^2 +6\chi^2 {\lt\langle 0 \rt|} \hat\psi^2 {\lt|0 \rt\rangle}+\chi^4.
\end{array}
\end{equation*}
Let us introduce the designation for the vacuum average of the
square of the over-condensate field operator:
\begin{equation} \label{GrindEQ__23_}
\rho \equiv {\lt\langle 0 \rt|} \hat\psi^2 {\lt| 0 \rt\rangle} .
\end{equation}
As is known, vacuum fluctuations of electromagnetic field lead to
observable physical effects, such, for example, as Lamb shift and
Casimir effect, and therefore must be taken into account in a
consistent quantum theory of any field. It is all the more important
in the case of states with broken symmetry, which can be
characterized by a nonzero vacuum average of the field operator
\eqref{GrindEQ__16_}. When taking account of the vacuum average of
the field operator, there is no reason to neglect the effects
conditioned by the vacuum average of the square of the field
operator. Let us note that taking account of the averages of the
field operator in powers higher than two, over the vacuum state of
the quadratic Hamiltonian \eqref{GrindEQ__15_}, does not have sense
because such averages are expressed in terms of the vacuum averages
\eqref{GrindEQ__16_}, \eqref{GrindEQ__23_}. It is also worth noting
that the non-operator term $\Omega$ in the Hamiltonian
\eqref{GrindEQ__15_} and in the Hamiltonian of interaction
\eqref{GrindEQ__10_} is quite substantial. The last of the
conditions \eqref{GrindEQ__19_}, that leads to the formula
\eqref{GrindEQ__22_}, ensures fulfillment of a natural condition
\begin{equation} \label{GrindEQ__24_}
{\lt\langle 0\rt|}H_C{\lt|0\rt\rangle}=0.
\end{equation}
Since the Hamiltonian $H_C$ describes the interaction of particles,
then, naturally, in the vacuum state the energy of this interaction
must be equal zero. Just taking into account of the non-operator
term $\Omega$ ensures fulfillment of this condition. Moreover, as
will be seen, the energy of the vacuum state is expressed in terms
of the quantity $\Omega$. From
\eqref{GrindEQ__21_}\,--\,\eqref{GrindEQ__23_} it follows that
\begin{equation} \label{GrindEQ__25_}
\Omega =-3g'\lt(\rho ^2 +2\rho \chi ^2 -\chi ^4 \rt).
\end{equation}
With the help of \eqref{GrindEQ__14_} and \eqref{GrindEQ__20_},
\eqref{GrindEQ__21_} we obtain the equation that determines the
order parameter $\chi$:
\begin{equation} \label{GrindEQ__26_}
\chi\lt(\kappa^2+12g'\rho+4g'\chi^2\rt)=0.
\end{equation}
This equation has two solutions. One of them, $\chi =0$, corresponds
to the phase with unbroken symmetry. The second one, with $\chi \ne
0$, is found from the requirement of equality to zero of the
expression in parentheses in \eqref{GrindEQ__26_}. We will assign
the index $s$ to the quantities relating to the symmetrical state
with $\chi =0$, and the index $b$ -- to those relating to the state
with broken symmetry, $\chi\ne 0$.
\\
\\
\noindent \textbf{PARTICLE MASSES. PHASE DIAGRAM}

The proposed approach lets us calculate the masses of particles,
\emph{i.e.} express them in terms of the parameters of the initial
Lagrangian \eqref{GrindEQ__1_}. From the Hamiltonian
\eqref{GrindEQ__15_} it follows that the parameter $D$ is connected
with the square of the particle mass:
\begin{equation}\label{GrindEQ__27_}
D=-\frac{m^2}{2}=-\frac{\kappa^2}{2}-6g'\lt(\rho+\chi^2\rt).
\end{equation}
As we see, particle masses in this approach are determined not only
by the mass parameter $\kappa^2$ that enters into initial the
Lagrangian \eqref{GrindEQ__1_} but also by the effects caused by
quantum fluctuations; moreover, they depend on the value of the
interaction constant. In symmetrical case the mass, according to
\eqref{GrindEQ__27_}, is defined by the relation
\begin{equation} \label{GrindEQ__28_}
m_{s}^2 =\kappa ^2 +12g'\rho_s,
\end{equation}
and the non-operator term \eqref{GrindEQ__25_} has the form
\begin{equation} \label{GrindEQ__29_}
\Omega_s=-3g'\rho_s^2.
\end{equation}
When the symmetry is broken, we have the following equation  for the
order parameter:
\begin{equation} \label{GrindEQ__30_}
\kappa^2+12g'\rho_b+4g'\chi^2=0.
\end{equation}
In this case the mass is determined by the value of the order
parameter
\begin{equation} \label{GrindEQ__31_}
m_b^2 =8g'\chi^2
\end{equation}
and $\Omega_{b}$\, by the formula \eqref{GrindEQ__25_}. Let us note
that, if we do not take into account the vacuum average of the
square of the field operator, by formally supposing $\rho=0$, then
from the formula \eqref{GrindEQ__30_} follows the relation
$\chi^2=-6\kappa^2/g$ obtained by Goldstone \cite{6}.

In the theory of non-relativistic many-particle systems with finite
density of number of particles, one understands by a self-consistent
field such mean field acting on a given particle, which is created
by all other particles of a considered system. In relativistic
theory, where the mean density of the number of particles equals
zero, the self-consistent field, as is seen from the relations
\eqref{GrindEQ__23_},\,\eqref{GrindEQ__27_},\, is formed by the
vacuum field fluctuations. By using the self-consistent approach for
determining the one-particle states, we actually account for the
influence of the vacuum fluctuations on the dynamical properties of
a particle.

The equation of motion for the over-condensate field operator
$\hat\psi$ in the  interaction picture has the form of the
Klein-Gordon equation
\begin{equation} \label{GrindEQ__32_}
\Delta \hat\psi\lt(x\rt)-\frac{\partial ^2 \hat\psi\lt(x\rt)}{\partial t^2} -m^2 \hat\psi\lt(x\rt)=0.
\end{equation}
Depending on whether in symmetrical or asymmetrical state a system
stays, one should understand by $m^2$ in \eqref{GrindEQ__32_} either
$m_s^2$\, or $m_b^2$. So, unlike Heisenberg field operator that
obeys the non-linear operator equation \eqref{GrindEQ__3_}, the
field operator in the interaction picture obeys the linear
Klein-Gordon equation with a positive, as will be shown below, value
of the mass square. The over-condensate field operator and also the
conjugate momentum can be presented in the form of expansion into
Fourier integral:
\begin{equation} \label{GrindEQ__33_}
\begin{array}{c}
{\hat\psi\lt(x\rt)=\lt(2\pi \rt)^{-3/2} \int d{\bf q}\, \, \lt(2q_{0} \rt)^{-1/2} \lt[a\lt(q\rt)e^{iqx} +a^{+} \lt(q\rt)e^{-iqx} \rt] {\kern 1pt} ,}
\\ \\
{\hat\pi\lt(x\rt)=-i\lt(2\pi \rt)^{-3/2} \int d{\bf q}\sqrt{q_{0}/2}
\lt[a\lt(q\rt)e^{iqx} -a^{+} \lt(q\rt)e^{-iqx} \rt],{\kern 1pt}  }
\end{array}
\end{equation}
where $q=\lt({\bf q},iq_{0} \rt)\, ,\quad q_{0} =\sqrt{m^2+{\bf
q}^2} $, $qx=q_{\mu}x_{\mu}={\bf qx}-q_0t$. Commutation relations of
operators at coinciding times have the usual for Bose-particles form
\[\lt[a\lt(q\rt),a\lt(q'\rt)\rt]=\lt[a^{+} \lt(q\rt),a^{+} \lt(q'\rt)\rt]=0\, ,\quad \quad \lt[a\lt(q\rt),a^{+} \lt(q'\rt)\rt]=\delta \lt(\textbf{q}-\textbf{q}'\rt). \] %
As far as the equation \eqref{GrindEQ__32_} is linear, we can
interpret operators $a^+\lt(q\rt)$ and $a\lt(q\rt)$ as the operators
of creation and annihilation of particles, and the state vector
$a^+\lt(q\rt){\lt| 0 \rt\rangle}$ as one that describes the
one-particle state with momentum $\textbf{q}$ and energy
$q_0=\sqrt{m^2+\textbf{q}^2}$. Using the representation
\eqref{GrindEQ__33_}, let us calculate the vacuum average
\eqref{GrindEQ__23_} that can be presented in the form
\begin{equation} \label{GrindEQ__34_}
\rho =\frac{1}{2\lt(2\pi\rt)^3} \int \frac{d{\bf q}}{\sqrt{{\bf q}^2+m^2}} =-\mathop{\lim }\limits_{\varepsilon \to +0} \frac{i}{\lt(2\pi \rt)^4 } \int \frac{dq}{q^2+m^2 -i\varepsilon }   .
\end{equation}
Here $dq\pp=\p d{\bf q} \pppp dq_0$, $\,q^2\!\p=\p{\bf q}^2-q_0^2$.
As is known, the integral in \eqref{GrindEQ__34_} diverges at large
momenta. It is accepted  \cite[\S 73]{18} that this divergence must
not cause any trouble, because the square of amplitude in any point
is not a measurable quantity. However, as we see, in this approach
the vacuum average of the square of the field operator is directly
related to an observed characteristic -- the particle mass. It seems
to be physically grounded, because spreading of a particle in the
vacuum occurs against a background of its fluctuations and,
consequently, the particle mass must substantially depend on the
intensity of such fluctuations. In order to provide the finiteness
of particle masses, there appears a need to introduce the cutoff
parameter $\Lambda$ at large momenta. Such parameter was introduced
also in earlier works \cite{4,5}, in which the methods of
superconductivity were firstly used for construction of models of
elementary particles. Let us note that in the self-consistent
equation of the superconductivity theory we also have to cut off the
integral at large momenta. Here, however, natural characteristic
scales are present, such as the average interparticle distance and
the effective radius of the interparticle potential. As is known, a
certain arbitrariness in the mode of regularization of divergent
integrals is present. Let us introduce the cutoff parameter in a
relativistically invariant way. When calculating the integral in
\eqref{GrindEQ__34_} we perform the Wick rotation of the integration
contour and use the substitution $q_0=iq_4$ \cite{9}. As a result we
obtain
\begin{equation} \label{GrindEQ__35_}
\rho=\Lambda^2 f\lt(\tilde{m}^2 \rt)\big/2\pi^2,
\qquad
f\lt(\tilde{m}^2 \rt)=1-\tilde{m}^2 \ln \lt(1+\tilde{m}^{-2}\rt),
\end{equation}
where $\tilde{m}^2={m^2 / \Lambda^2 } $. The function
$f\lt(\tilde{m}^2 \rt)$ is equal to unity at $\tilde{m}^2=0$, and
monotonically decreases tending to zero at $\tilde{m}^2 \to \infty$.
By substituting the expression for the vacuum average
\eqref{GrindEQ__35_} into \eqref{GrindEQ__28_} and
\eqref{GrindEQ__31_}, we find the self-consistent equations that
define the square of mass in symmetrical (a) and asymmetrical (b)
phases:
\begin{subequations}\label{GrindEQ__36_}
\begin{align}
\qquad \tilde{m}_s^2 -\tilde{\kappa }^2 =g''\lt[1-\tilde{m}_s^2 \ln \lt(1+\tilde{m}_s^{-2} \rt)\rt],\label{eq_36a}
\\
\quad -\frac{\tilde{m}_b^2 }{2} -\tilde{\kappa }^2 =g''\lt[1-\tilde{m}_b^2 \ln \lt(1+\tilde{m}_b^{-2} \rt)\rt],\label{eq_36b}
\end{align}
\end{subequations}
where $\tilde{m}_i^2 =m_{i}^2\big/\Lambda ^2 \, \, (i=s,b)$,
$\tilde{\kappa}^2 =\kappa^2\big/\Lambda^2$,  $g''=6g'/\pi=g/4\pi$.
The equations \eqref{GrindEQ__36_}  define the square of mass
related to the square of the cutoff parameter as a function of the
dimensionless mass parameter $\tilde{\kappa }^2$ and the interaction
constant $g''$: $\tilde{m}_{i}^2 =\tilde{m}_i^2 \lt(\tilde{\kappa}^2
,g''\rt)$. It is noteworthy that, while the particle mass
substantially depends on the cutoff parameter, the equations
\eqref{GrindEQ__36_} for the dimensionless masses $\tilde{m}_i^2$ do
not already contain it explicitly.

\begin{figure}
\centering
\includegraphics[width = 0.7\columnwidth]{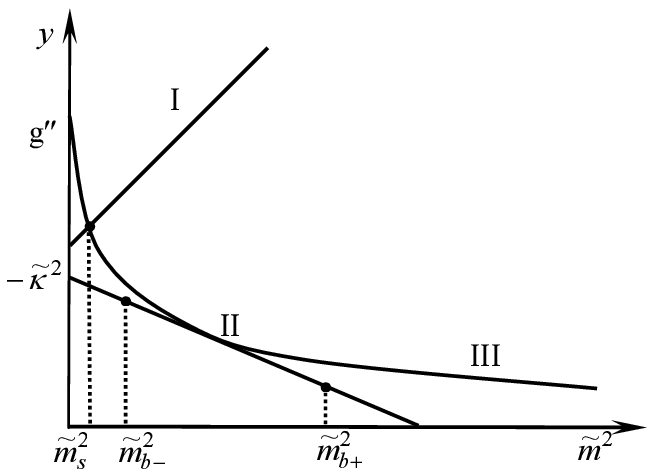}\\
Figure~1: Graphical solution of the equations \eqref{GrindEQ__36_}.\\
\vspace{-0.7em} %
\begin{flushleft}
\hspace{2cm} I: $y=\tilde{m}^2 -\tilde{\kappa}^2$,
\,\, II: $y=-\tilde{m}^2/2-\tilde{\kappa}^2$, 
\,\, III: $y=g''\lt[1-\tilde{m}^2 \ln \lt(1+\tilde{m}^{-2}\rt)\rt].$
\vspace{-1em} %
\end{flushleft}
\end{figure}

It is convenient to analyze these equations graphically (see
Figure~1). The right parts of the equations \eqref{GrindEQ__36_} are
identical, being a function that decreases monotonically from the
value $g''$ at $\tilde{m}^2 =0$ to zero at $\tilde{m}^2 \to \infty $
(curve III). Their left parts are functions linear in the variable
$\tilde{m}^2$, which branch off from the point $-\tilde{\kappa }^2
$. The point of intersection of the straight line $y=\tilde{m}^2
-\tilde{\kappa }^2$ (curve I) with the curve III gives the value of
the particle mass in symmetrical phase (s-phase) $\tilde{m}_{s}^2 $,
and the intersection of the line $y=-{\tilde{m}^2/2}
-\tilde{\kappa}^2$ (curve II) with the curve III gives the value of
the particle mass in asymmetrical phase (b-phase) $\tilde{m}_{b}^2
$. It is obvious from Figure~1 that, depending on the value of the
parameter $-\tilde{\kappa }^2$, for the s-phase there exists either
one solution or none. For the b-phase the solutions are either
absent or there exist one or two solutions. Here we can mark out two
particular cases. The first one takes place when the condition
$\tilde{\kappa }^2 =-g''$ holds. Then there exist one solution for
s-phase with zero mass, and two solutions for b-phase. One of them
also corresponds to zero mass, and the other one to a finite mass
\begin{equation} \label{GrindEQ__37_}
\tilde{m}_{b**}^2 =\lt(e^{1/2g''}-1\rt)^{-1}.
\end{equation}
The second particular solution corresponds to the case when the
straight line $y=-{\tilde{m}^2/2} -\tilde{\kappa }^2$ is a tangent
to the curve $y=g''f(\tilde{m}^2 )$. The solution of the equation
(36b) in the tangency point is determined by the equation
\begin{equation} \label{GrindEQ__38_}
\lt(2g''\rt)^{-1}=\ln\lt(1+\tilde{m}_{b*}^{-2}\rt)-\lt(1+\tilde{m}_{b*}^2\rt)^{-1},
\end{equation}
and corresponding value of the mass parameter is given by the
formula
\begin{equation} \label{GrindEQ__39_}
\tilde{\kappa }_{*}^2 =-g''\mmm\Big/\mmm\lt(1+\tilde{m}_{b*}^2 \rt).
\end{equation}
The particle mass in the symmetrical phase $\tilde{m}_{s*}^2 $ at
the value of the mass parameter \eqref{GrindEQ__39_} is defined by
the equation \eqref{eq_36a}, where it is necessary to assume
$\tilde{\kappa }^2 =\tilde{\kappa }_{*}^2$. So, the following
solutions of the equations \eqref{GrindEQ__36_} are possible:

a) $\tilde{\kappa }_{*}^2 <\tilde{\kappa }^2 <\infty $: there exists
a solution for s-phase only ($\tilde{m}_s^2 >0$); the solutions for
b-phase are absent;

b) $\tilde{\kappa }_{*}^2 =\tilde{\kappa }^2 $: there exists a
solution for s-phase ($\tilde{m}_{s*}^2 >0$) and appears one
solution for b-phase ($\tilde{m}_{b*}^2>\tilde{m}_{s*}^2$);

c) $-g''<\tilde{\kappa }^2 <\tilde{\kappa }_{*}^2 $: there exist a
solution for s-phase ($\tilde{m}_{s}^2 >0$) and two solutions for
b-phase ($\tilde{m}_{b+}^2 >\tilde{m}_{b-}^2 >\tilde{m}_{s}^2 $),
let us call the one with $\tilde{m}_{b+}^2$ a b$_{+}$-phase, and the
one with $\tilde{m}_{b-}^2$ $-$ a b$_{-}$-phase;

d) $\tilde{\kappa }_{*}^2 =-g''$: there exist the solution
$\tilde{m}_{s}^2 =0$ for s-phase and two solutions for b-phase, the
first one $\tilde{m}_{b}^2 =0$, and the second one
$\tilde{m}_{b**}^2 >0$ \eqref{GrindEQ__37_};

e) $-\infty <\tilde{\kappa }^2 <-g''$: the solutoin for s-phase is
absent and there exists the only solution $\tilde{m}_{b}^2>0$ for
b-phase.

\begin{figure}
\centering
\includegraphics[width = 0.7\columnwidth]{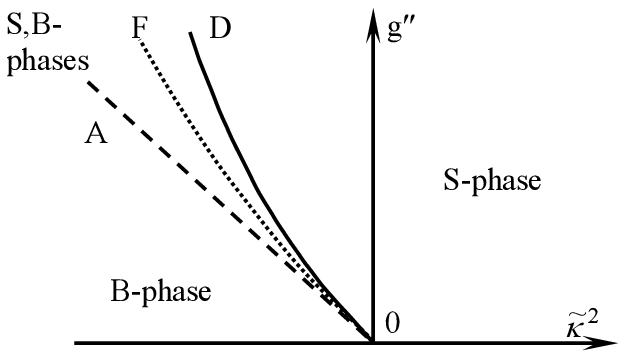}\\
Figure~2: Areas of existence of phases of a real scalar field.
\end{figure}

It is convenient to present the ranges of parameters, where
different solutions exist, on the phase diagram (see Figure~2). The
ordinates axis is the value of the interaction constant $g''$ and
the absciss axis is the value of the mass parameter $\tilde{\kappa
}^2$. In the area S lying rightwards from the curve OD, which is
given by the equations \eqref{GrindEQ__38_},\,\eqref{GrindEQ__39_},
there exists the symmetrical phase only. In the area B lying
leftwards from the curve OA ($g''=-\kappa ^2$), there exists the
asymmetrical phase only. In the area between the curves OA and OD,
the existence of both s-phase and two b-phases with different values
of the particle masses is possible. In order to determine which of
three possible phases is stable in the indicated area, it is
necessary to compare the energy densities of their vacuums. The
Hamiltonian \eqref{GrindEQ__15_}, with account of the expansions
\eqref{GrindEQ__33_}, has the form
\begin{equation} \label{GrindEQ__40_}
H_{0} =\int d\textbf{q} \, q_{0} \, a^{+} (q)a(q)+C+V\lt(-4g'\chi ^4 +\Omega \rt),
\end{equation}
where the constant
\begin{equation} \label{GrindEQ__41_}
C=\frac{1}{2}\int d\textbf{q}d\textbf{q}'\p q_{0}\pp\delta \lt(\textbf{q}-\textbf{q}'\rt)\delta \lt(\textbf{q}-\textbf{q}'\rt)
\end{equation}
appears when passing to a normal order of operators in the
Hamiltonian $H_0$. So, from \eqref{GrindEQ__40_} it follows that the
ground state energy of a scalar field (the vacuum) is determined by
the formula $E_V=E_0+C$, where
\begin{equation} \label{GrindEQ__42_}
E_{0} =V\lt(-4g'\chi ^4 +\Omega \rt).
\end{equation}
The energy $E_0$ is different in s- and b-phases. With regard to the
formulas \eqref{GrindEQ__25_},\,\eqref{GrindEQ__29_} we find for
s-phase
\begin{equation} \label{GrindEQ__43_}
\frac{E_{0s}}{V} =-\frac{1}{g''}\pp\varepsilon_{0} \lt(\tilde{m}_{s}^2 -\tilde{\kappa }^2 \rt)^2 ,
\end{equation}
and for b-phase
\begin{equation} \label{GrindEQ__44_}
\frac{E_{0b} }{V} =-\frac{1}{g''}\ppp\varepsilon_{0} \lt[\lt(\tilde{m}_{b}^2 -\tilde{\kappa }^2 \rt)^2 -{3\tilde{m}_{b}^4/2} \rt],
\end{equation}
where $\varepsilon_{0} ={\Lambda ^4\mm/8\pi ^2 } $. The constant $C$
is infinite, an with regard to the fact that $\delta
\lt(0\rt)={V/\mm\lt(2\pi \rt)^3 } $, it can be presented in the form
\begin{equation} \label{GrindEQ__45_}
C=\frac{V}{2\lt(2\pi \rt)^3 } \int d\textbf{q} \sqrt{\textbf{q}^2+m^2} =\frac{V\, J}{2\lt(2\pi \rt)^3 },
\end{equation}
where $J\equiv\int d\textbf{q}\sqrt{\textbf{q}^2+m^2}$. Since above
there was introduced a cutoff at large momenta, the integral $J$ can
be calculated if its regularization is  carried out by the same way
as in calculating the integral of the vacuum average in
\eqref{GrindEQ__34_}. By differentiating $J$ with respect to $m^2$,
we come to the obtained earlier integral \eqref{GrindEQ__35_}, so
that
\begin{equation} \label{GrindEQ__46_}
\frac{dJ}{d\pp m^2 } =\frac{1}{2} \int d\textbf{q}\lt(\textbf{q}^2 +m^2 \rt)^{-1/2}=4\pi \Lambda ^2 \lt[1-\tilde{m}^2 \ln \lt(1+\tilde{m}^{-2} \rt)\rt].
\end{equation}
By integrating the last relation we obtain
\begin{equation} \label{GrindEQ__47_}
C=V\varepsilon_{0} \lt[-\tilde{m}^4 \ln \lt(1+\tilde{m}^{-2} \rt)+\ln \lt(1+\tilde{m}^2 \rt)+\tilde{m}^2 +c'\rt],
\end{equation}
where $c'$ is an integration constant, which is not dependent on the
system parameters, and because of this fact can be put equal to
zero. In this case the parameter $C$ is always positive and raises
monotonically with the increase of $\tilde{m}^2$. So, the vacuum
energy is determined by two contributions. The contribution defined
by the parameter $C$ \eqref{GrindEQ__47_} is conditioned by the
necessity of normal ordering of operators in the free Hamiltonian
$H_0$. This contribution into the vacuum energy is also present in
the model of free fields and is always positive for Bose fields. The
contribution of the term $E_0$ \eqref{GrindEQ__42_} is essentially
determined by taking account of nonlinearity at the step of
construction of the main approximation and, as one can see, gives a
negative contribution into the vacuum energy. With regard to the
equations \eqref{GrindEQ__36_} one can present the full density of
the vacuum energy in the form
\begin{equation} \label{GrindEQ__48_}
\frac{E_{V} }{V} =\varepsilon_{0} \lt[\frac{\kappa ^2 }{g''} \lt(\tilde{m}^2 -\tilde{\kappa }^2 \rt)+\ln \lt(1+\tilde{m}^2 \rt)\rt].
\end{equation}
The formula \eqref{GrindEQ__48_} relates both to symmetrical phase
and to asymmetrical phases. Given that the particle mass is
determined by the parameters of Lagrangian, formula
\eqref{GrindEQ__48_} together with equations \eqref{GrindEQ__36_}
determines the vacuum energy as a function of Lagrangian parameters
$\kappa^2$ and $g$. Also, the vacuum energy substantially depends on
the cutoff parameter $\Lambda$. In Figure~3 the dependence of the
vacuum energy on the mass parameter is presented. The vacuum energy
is positive at large positive values of the mass parameter and
decreases with decrease of $\tilde\kappa^2$. Under some negative
value of $\tilde\kappa_0^2$ the vacuum energy vanishes, and becomes
negative at $\tilde\kappa^2 <\tilde\kappa_0^2$. In the range
$-g''<\tilde\kappa^2 <\tilde\kappa_*^2$ the existence of three
phases is possible. They are symmetrical s-phase and two phases with
broken symmetry b${}_{+}$ and b${}_{-}$ that correspond to higher
and lower masses. That one among them will be stable, which
corresponds to the minimal vacuum energy. In the range
$\tilde{\kappa }_{f}^2 <\tilde{\kappa }^2 <\tilde{\kappa}_*^2 $ the
minimal is the energy of s-phase. In the point $\tilde{\kappa }^2
=\tilde{\kappa }_{f}^2 $ (curve OF in Figure~2) a continuous phase
transition occurs without a jump of energy, and under $\tilde{\kappa
}^2 <\tilde{\kappa}_f^2$ stable is the phase with broken symmetry
with a larger value of mass (b$_+$-phase). %
\vspace{10mm}

\begin{figure}
\centering
\includegraphics[width=0.7\columnwidth]{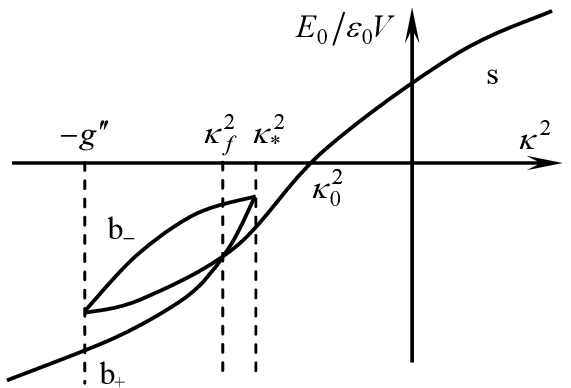}\\
Figure~3: Dependence of the vacuum energy density of phases \\ of a
real scalar field on the square of the mass parameter. %
\end{figure}

\vspace{5mm} %
\noindent \\ \textbf{PERTURBATION THEORY}

While constructing the perturbation theory, one frequently
postulates that the interaction Hamiltonian must be written in the
form of a normal product \cite{19}. In this way zero energy is at
once excluded from the theory. However, as was noted above, the
effects caused by zero oscillations of field can lead to observable
effects and substantially influence the particle dynamics. Notable
is the fact that, under the self-consistent mode of introduction of
the one-particle states that is described in the present work, the
interaction Hamiltonian \eqref{GrindEQ__10_}, without supplementary
suppositions about normal form of the initial Hamiltonian, takes the
form of a normal product of the field operators:
\begin{equation} \label{GrindEQ__49_}
\hat H_C \lt(x\rt)=\frac{g}{4!} N\big(\hat\psi^4
(x)\big)+\frac{g}{3!}\,\chi N\big(\hat\psi^3 (x)\big),
\end{equation}
where normal products of the field operators taken in one point can
be presented in the form $N\big(\hat\psi^4 (x)\big)=\hat\psi^4
(x)-6\, \hat\psi^2 (x)\rho +3\rho^2$,
$N\big(\hat\psi^3(x)\big)= \hat\psi^3 (x)-3\,\hat\psi(x)\rho$. %
The averages over the vacuum state of the normal products introduced
here are equal to zero: ${\lt\langle 0 \rt|} N\big(\hat\psi^4
(x)\big){\lt| 0 \rt\rangle} ={\lt\langle 0 \rt|} N\big(\hat\psi^3
(x)\big){\lt| 0 \rt\rangle} =0$, so that the condition ${\lt\langle
0 \rt|} \hat H_C (x){\lt| 0 \rt\rangle} =0$ is, obviously,
satisfied. Thus, in this approach the normal form of the interaction
Hamiltonian is not postulated \emph{ab initio} but arises as a
consequence of the choice of the self-consistent field model as the
main approximation, where the effects of zero fluctuations are
already taken into account. As far as the interaction Hamiltonian
\eqref{GrindEQ__49_} is a sum of two Hamiltonians, then the
contribution of the $n$th order into $S$-matrix
$S=\sum_{n=0}^{\infty}S^{\lt(n\rt)}$ can be written in the form
\newpage
\begin{equation} \label{GrindEQ__50_}
\begin{array}{ll}
\displaystyle{S^{\lt(n\rt)}=\dfrac{\lt(-i\rt)^n}{n\,!}
\int dx_1 \ldots dx_n\, T\bigg[\hat H_C^{\lt(4\rt)} \lt(x_1 \rt)\ldots \hat H_C^{\lt(4\rt)} \lt(x_n \rt) +\hat H_C^{\lt(3\rt)} \lt(x_1 \rt)\ldots \hat H_C^{\lt(3\rt)} \lt(x_n \rt)+}\\
\displaystyle{\hspace{10mm} + \sum_{m=1}^{n-1}C_n^m\, \hat
H_C^{\lt(4\rt)} \lt(x_1 \rt)\ldots \hat H_C^{\lt(4\rt)} \lt(x_m
\rt)\hat H_C^{\lt(3\rt)} \lt(x_{m+1} \rt)\ldots \hat H_C^{\lt(3\rt)}
\lt(x_n \rt) \bigg]\, ,}
\end{array}
\end{equation}
where $\hat H_C^{\lt(4\rt)} (x)=\dfrac{g}{4!} N\big(\hat\psi^4
(x)\big)$, $H_C^{\lt(3\rt)}(x)=\dfrac{g}{3!} \pp\chi \,
N\big(\hat\psi^3 (x)\big)$, $C_n^m$ -- binomial coefficients, $T$ --
chronological operator. Perturbation theory is constructed in the
standard way. Feinman diagrams in this case contain the elements
presented in Figure~4. The solid line outgoing from the point $x$,
which describes the creation of a particle with 4-momentum $q$,
corresponds to the expression $f_q\lt(x\rt)={e^{-iqx} \mathord{\lt/
{\vphantom {e^{-iqx}  \lt(2\pi \rt)^{3/2} \sqrt{2q_0} }} \rt.
\kern-\nulldelimiterspace} \lt(2\pi \rt)^{3/2} \sqrt{2q_{0} } } $
(Figure~4a), and the solid line incoming into the point $x$, which
describes the annihilation of a particle with 4-momentum $q$,
corresponds to the expression $f_q^*\lt(x\rt)={e^{iqx} \mathord{\lt/
{\vphantom {e^{iqx}  \lt(2\pi \rt)^{3/2} \sqrt{2q_{0} } }} \rt.
\kern-\nulldelimiterspace} \lt(2\pi \rt)^{3/2} \sqrt{2q_{0} } } $
(Figure~4b). The solid line connecting the points $x_1$ and $x_2$
(Figure~4c) corresponds to the expression $-G^{\lt(0\rt)}
\lt(x_1-x_2\rt)$, where Green function has the form
\begin{equation} \label{GrindEQ__51_}
G^{\lt(0\rt)} \lt(x\rt)=\mathop{\lim }\limits_{\varepsilon \to +0} \frac{i}{\lt(2\pi \rt)^4 } \int \frac{dq\, e^{iqx} }{q^2 +m^2 -i\varepsilon }  .
\end{equation}
Let us note that, according to \eqref{GrindEQ__34_} and
\eqref{GrindEQ__51_}, $G^{(0)}(0)=-\rho$. The condensate function
$i\chi$ is put in correspondence to a wavy line with a cross
(Figure~4d), and the vertex corresponds to a factor $-ig$
(Figure~4e). There are two types of elementary vertices. Either four
solid lines or three solid and one wavy line can converge in a
vertex. If a permutation of $n$ inner lines does not change the form
of the diagram, then when writing a matrix element one should use
the factor $1/n!$. As far as the interaction Hamiltonian in the
proposed variant of perturbation theory has the normal form
\eqref{GrindEQ__49_}, then the diagrams that contain ``loops'',
\emph{i.e.} solid lines whose ends converge in the same vertex, are
absent (Figure~4f). Since the factor $\rho$ is matched to a loop,
then it is obvious that the diagrams with loops were in fact taken
into account when constructing the main approximation and
calculating the particle mass. The circumstance that the interaction
$g$ is already to a certain extent accounted for when constructing
the main approximation and the one-particle states, leads to that
the perturbation theory formulated in this way remains correct also
in the case when the interaction constant is not small. Thus, the
proposed method of construction of the perturbation theory can prove
to be effective for description of strongly interacting particles.
The fact that the area of applicability of the perturbation theory
constructed on the basis of the presented approach is much larger
than the area of applicability of the standard theory is shown on
the example of anharmonic oscillator \cite{20}.

It is of interest to compare the proposed variant of the
perturbation theory with the perturbation theory constructed on the
basis of the Goldstone approach \cite{6}. The free particle
Lagrangian in the Goldstone model is obtained by carrying out a
shift of the field by the value which is found from the condition of
minimum of the potential energy of a classical scalar field
\cite{6,8}. The Goldstone result can also be obtained by another
way, directly from considering a quantized field. For this, in the
Lagrangian \eqref{GrindEQ__1_} we will go over to a new field by
performing the substitution
$\hat\varphi\lt(x\rt)=\hat\psi\lt(x\rt)+\chi $ and require that the
coefficient at appeared linear in the field $\hat\psi\lt(x\rt)$ term
turns into zero. This leads to the equation that determines the
parameter $\chi$:
\begin{equation} \label{GrindEQ__52_}
\chi \lt(\kappa^2 +\frac{g}{6} \chi^2 \rt)=0.
\end{equation}
\begin{figure}
\centering
\includegraphics[width = 0.7\columnwidth]{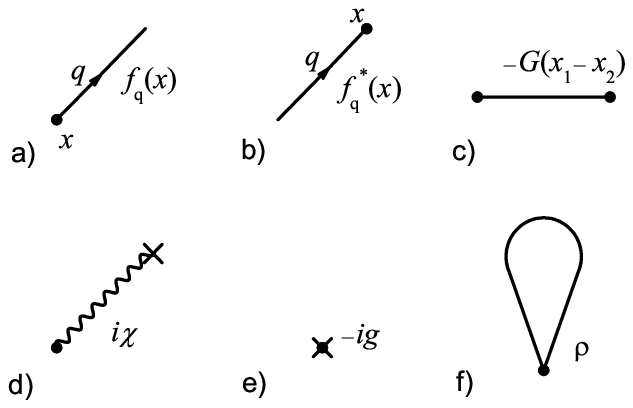}\\
Figure~4: Elements of Feinman diagrams of a real scalar field with broken symmetry.
\end{figure}
The equation \eqref{GrindEQ__52_} differs from \eqref{GrindEQ__27_}
by the circumstance that the expression in parentheses does not
contain a term with the vacuum average of the square of the
over-condensate field operator $\rho $, \emph{i.e.} the effects
caused by vacuum fluctuations are not taken into account. At
$\kappa^2>0$ (remember that $g>0$) the equation \eqref{GrindEQ__52_}
has the only solution $\chi=0$. At $\kappa^2<0$ the solution
$\chi=0$ does not describe real particles, since the square of their
mass is negative. The particles with a correct sign of the square of
mass are described by the second solution of \eqref{GrindEQ__52_},
$\chi^2=-6\kappa^2/g$, that coincides with the result obtained in an
other approach by Goldstone \cite{6}, and the square of mass of such
particles is $m^2=-2\kappa^2$. Let us remark that here the masses of
free particles are determined by only the mass parameter and do not
depend on both vacuum fluctuations and the interaction constant. The
interaction Hamiltonian in this case coincides with the expression
\eqref{GrindEQ__49_}, if the normal product sign is dropped there,
and consequently we must take into account the diagrams containing
loops in all orders of the perturbation theory. Thus, the
construction of the perturbation theory on the basis of the choice
of the self-consistent field model as the main approximation lets us
significantly simplify the perturbation theory, because a very large
number of diagrams with loops are already taken into account in the
zero approximation.

\noindent \\ \textbf{QUASI-AVERAGES. EXACT GREEN FUNCTIONS}

Since the zero approximation Hamiltonian \eqref{GrindEQ__9_}
contains the terms linear in the field operators
$\hat\varphi\lt(x\rt)$, so that its symmetry relative to the
operation $\varphi \to -\varphi $ is broken, there follows the
distinction from zero of the vacuum average of the field operator,
${\lt\langle 0 \rt|}\hat\varphi{\lt|0\rt\rangle} = \chi \ne 0$. The
interaction Hamiltonian \eqref{GrindEQ__10_} is not invariant under
the operation $\varphi \to -\varphi$ too; however, the full
Hamiltonian $H=H_0+H_C$ remains invariant to this operation. It is
easy to show that the average over the vacuum state of a system with
the invariant Hamiltonian $H$ equals zero. Indeed, let $\lt|\lt.
0\rt)\rt. $ be the vacuum vector of the system $H$, so that
$H\lt|\lt.0\rt)=E_0\lt|\lt.0\rt)\rt.\rt.$, where $E_0$ is the vacuum
energy. Also, let Hamiltonian be invariant under some unitary
transformation $UHU^+=H$. Then, obviously, $U\lt|\lt. 0\rt)\rt.$ is
also the state vector corresponding to the energy $E_0$. We consider
the vacuum state with the energy $E_0$ as unique, \emph{i.e.}
$U\lt|\lt. 0\rt)\rt.$ coincides with $\lt|\lt.0\rt)\rt.$:
$U\lt|\lt.0\rt)=e^{i\alpha}\lt|\lt.0\rt)\rt.\rt.$ except for a phase
factor. In the considered case of a scalar field with broken
symmetry the unitary transformation is such that $U\varphi\pp
U^{+}=-\varphi$. In consequence of invariance of the Hamiltonian and
uniqueness of the vacuum we have
$\lt(0\lt|\pp\varphi\rt|0\rt)=-\lt(0\lt|\pp U^+\varphi\pp
U\rt|0\rt)=-\lt(0\lt|\varphi\rt|0\rt)=0$. So, in order to obtain a
state with non-zero average, $\lt(\lt. 0\rt|\varphi \lt|\lt.
0\rt)\ne \rt. \rt. 0$, one should break the symmetry of the initial
Hamiltonian. Such an approach to the description of states with
broken symmetry in statistical mechanics in the framework of the
concept of quasi-averages was developed by Bogoljubov \cite{16}. An
idea to use a model of self-consistent field for determining
quasi-averages was proposed in \cite{21}. Let us agree to use for
the description of systems with broken symmetries a Hamiltonian of a
more common than $H=H_0+H_C$ form, namely $H_{\nu}=H_{0}+\nu\,H_C$,
where $\nu$ is a real positive parameter. If $\nu =0$, then $H_{\nu}
$ coincides with the self-consistent field Hamiltonian, but if $\nu
=1$, then $H_{\nu}$ coincides with the full Hamiltonian $H$
\eqref{GrindEQ__5_}. By changing the parameter $\nu$ from zero to
unity, we ``switch on'' the interaction between the particles
defined in the self-consistent field model. If the parameter $\nu$
is not equal to unity, but is close to it as is wished, then the
Hamiltonian $H_\nu$ is as wished close to the Hamiltonian $H$.
However, there is a principal difference between these Hamiltonians
in the situation when the symmetry is violated \emph{i.e.} the
symmetry of $H_0$ is lower than the symmetry of the initial
Hamiltonian $H$. In this case, at $\nu \ne 1$ the Hamiltonian
$H_{\nu}$ contains an additional term that lowers its symmetry to
the symmetry of the Hamiltonian $H_0$, that allows us to describe
the states with the symmetry lower than that of $H$. The ground
state vector, determined by the Schr\"odinger equation
$H_\nu\lt|\lt.0_{\nu}\rt) = E_{0\nu}\lt|\lt.0_{\nu}\rt)\rt.\rt.$, as
well as the vacuum energy and also the field operators in Heisenberg
picture depend on the parameter $\nu$. We will define the
quasi-average over the vacuum state of the operator $A$ by the
expression
\begin{equation} \label{GrindEQ__53_}
\lt\langle A\rt\rangle_{0} \equiv \mathop{\lim }\limits_{\nu \to 1} \lt(\lt. 0_{\nu} \rt|\rt. A\lt|\lt. 0_{\nu } \rt)\rt. .
\end{equation}
The passage to the limit $\nu \to 1$ must be performed after the
passage $V\to\infty$ is done. The limit in the right side of the
definition \eqref{GrindEQ__53_}, generally speaking, may not
coincide with the average $\lt(0\rt|\p A\p\lt|0\rt)$ calculated over
the system vacuum state with the Hamiltonian $H_{\nu}$ at $\nu=1$.
So, the quasi-average \eqref{GrindEQ__53_} can be distinct from
zero, even if, by virtue of the symmetry properties of the
Hamiltonian $H$, the average $\lt(0\rt|\p A\p\lt|0\rt)$ is equal to
zero.

Now let us give the equations for exact Green functions of a real
scalar field. The exact \textit{n}-point over-condensate Green
function is determined by the relation
\begin{equation} \label{GrindEQ__54_}
G^{\lt(n\rt)} \lt(x_1 \ldots x_{n} \rt)=i^{n} \lt\langle T\lt[\psi \lt(x_1 \rt)\ldots \psi \lt(x_{n} \rt)\rt]\rt\rangle_{0},
\end{equation}
where we should understand averaging in the sense of quasi-average
\eqref{GrindEQ__53_}. Let us note that for the states with broken
symmetries the Green functions with odd $n$ can be distinct from
zero, too. We write the full Hamiltonian in the form of a sum of
self-consistent and correlation Hamiltonians in terms of operators
in Heisenberg picture. By using the equations of motion for the
field operator and the conjugated momentum, and also the commutation
relations, we obtain the following equations for the one-point and
two-point Green functions:
\begin{equation} \label{GrindEQ__55_}
\lt(\frac{\partial^{\pp2}}{\partial x^2 } -m^2 +\frac{g}{2} \rho \rt)\, G^{\lt(1\rt)} \lt(x\rt)+\frac{g}{2} \, i\chi \lt(G^{\lt(2\rt)} \lt(x\rt)+\rho \rt)+\frac{g}{3!} G^{\lt(3\rt)} \lt(x\rt)=0,
\end{equation}
\begin{equation} \label{GrindEQ__56_}
\begin{array}{l}
{\lt(\dfrac{\partial^{\pp2}}{\partial x_1^2}-m^2+\dfrac{g}{2} \rho \rt)\, G^{\lt(2\rt)} \lt(x_1 x_2 \rt)-i\dfrac{g}{2} \chi \lt(G^{\lt(3\rt)} \lt(x_1 x_1 x_2 \rt)+\rho \, G^{\lt(1\rt)} \lt(x_2 \rt)\rt)+}
\\
{\hspace{62mm}+\dfrac{g}{3!} G^{\lt(4\rt)} \lt(x_1 x_1 x_1 x_2 \rt)=-i\delta \lt(x_1 -x_2 \rt),} %
\end{array}
\end{equation}
where Green functions at coinciding arguments are determined by the
relation $G^{\lt(n\rt)} \lt(x\rt)\equiv \lt\langle \psi ^{n}
\lt(x\rt)\rt\rangle_{0} $. By introducing the vertex functions
$\Gamma ^{\lt(3\rt)} \lt(x_1 x_2 x_3 \rt), \Gamma ^{\lt(4\rt)}
\lt(x_1 x_2 x_3 x_4 \rt)$, we can present the three- and four-point
Green functions in another form:
\begin{equation} \label{GrindEQ__57_}
\begin{array}{c}
{G^{\lt(3\rt)} \lt(x_1 x_2 x_3 \rt)=G^{\lt(2\rt)} \lt(x_1 x_2
\rt)G^{\lt(1\rt)} \lt(x_3 \rt)+G^{\lt(2\rt)}}\lt(x_1 x_3
\rt)G^{\lt(1\rt)} \lt(x_2 \rt)+ G^{\lt(2\rt)} \lt(x_2 x_3
\rt)G^{\lt(1\rt)} \lt(x_1 \rt)+
\vspace{1mm}\\
{\hspace{13mm}+\int dx'_1  dx'_2 dx'_3 \, \Gamma ^{\lt(3\rt)}
\lt(x'_1 x'_2 x'_3 \rt)G^{\lt(2\rt)} \lt(x_1 x'_1 \rt)G^{\lt(2\rt)}
\lt(x_2 x'_2 \rt)G^{\lt(2\rt)} \lt(x_3 x'_3 \rt) ,}
\end{array}
\end{equation}
\begin{equation} \label{GrindEQ__58_}
\begin{array}{lcc}
{G^{\lt(4\rt)} \lt(x_1 x_2 x_3 x_4 \rt)=}\vspace{1mm}\\ %
{\hspace{5mm}=G^{\lt(2\rt)} \lt(x_1 x_2 \rt)G^{\lt(2\rt)} \lt(x_3 x_4 \rt)+} %
{G^{\lt(2\rt)} \lt(x_1 x_3 \rt)G^{\lt(2\rt)} \lt(x_2 x_4 \rt)+G^{\lt(2\rt)} \lt(x_1 x_4 \rt)G^{\lt(2\rt)} \lt(x_2 x_3 \rt)+} %
\vspace{1mm}\\
\hspace{5mm}+\int dx'_1  dx'_2 dx'_3 dx'_4 \, \Gamma ^{\lt(4\rt)}
\lt(x'_1 x'_2 x'_3 x'_4 \rt)G^{\lt(2\rt)} \lt(x_1 x'_1
\rt)G^{\lt(2\rt)} \lt(x_2x'_2\rt)G^{\lt(2\rt)} \lt(x_3 x'_3
\rt)G^{\lt(2\rt)} \lt(x_4 x'_4 \rt) .
\end{array}
\end{equation}
With account of the last relations, \eqref{GrindEQ__55_} and
\eqref{GrindEQ__56_} take the following form:
\begin{equation} \label{GrindEQ__59_}
\begin{array}{c}
{\lt[\dfrac{\partial^{\pp2}}{\partial x^2} -m^2 +\dfrac{g}{2} \lt(G^{\lt(2\rt)} \lt(x\rt)+\rho \rt)\rt]G^{\lt(1\rt)} \lt(x\rt)+} %
{\,i\dfrac{g}{2} \chi \lt(G^{\lt(2\rt)} \lt(x\rt)+\rho \rt)+}
\vspace{1mm}\\
{\hspace{7mm}+\dfrac{g}{3!} \int dx'_1  dx'_2 dx'_3 \, \Gamma
^{\lt(3\rt)} \lt(x'_1 x'_2 x'_3 \rt)G^{\lt(2\rt)} \lt(xx'_1
\rt)G^{\lt(2\rt)} \lt(xx'_2 \rt)G^{\lt(2\rt)} \lt(xx'_3 \rt)=0,}
\end{array}
\end{equation}
\begin{equation} \label{GrindEQ__60_}
\begin{array}{c}
{\lt[\dfrac{\partial^{\pp2}}{\partial x_1^2}-m^2 +\dfrac{g}{2} \lt(G^{\lt(2\rt)} \lt(x_1 \rt)+\rho \rt)\rt]G^{\lt(2\rt)} \lt(x_1 x_2 \rt)-}
\vspace{1mm}\\
{-i\dfrac{g}{2} \chi \lt[G^{\lt(2\rt)} \lt(x_1 \rt)+\rho
\rt]G^{\lt(1\rt)} \lt(x_2 \rt)-ig\chi G^{\lt(2\rt)} \lt(x_1 x_2
\rt)G^{\lt(1\rt)} \lt(x_1 \rt)-}
\vspace{1mm}\\
{-i\dfrac{g\chi }{2} \int dx'_1 dx'_2 dx'_3 \, \Gamma ^{\lt(3\rt)}
\lt(x'_1 x'_2 x'_3 \rt)G^{\lt(2\rt)} \lt(x_1 x'_1 \rt)G^{\lt(2\rt)}
\lt(x_1 x'_2 \rt)G^{\lt(2\rt)} \lt(x_2 x'_3 \rt)+}
\vspace{1mm}\\
{+\dfrac{g}{3!} \int dx'_1  dx'_2 dx'_3 dx'_4 \, \Gamma ^{\lt(4\rt)}
\lt(x'_1 x'_2 x'_3 x'_4 \rt)G^{\lt(2\rt)} \lt(x_1 x'_1
\rt)G^{\lt(2\rt)} \lt(x_1 x'_2 \rt)G^{\lt(2\rt)} \lt(x_1 x'_3
\rt)G^{\lt(2\rt)} \lt(x_2 x'_4 \rt)=}
\vspace{1mm}\\
{=-i\delta \lt(x_1-x_2\rt).}
\end{array}
\end{equation}
The equations for Bose fields under broken symmetry have a more
complex structure than those for Fermi fields, owing to the presence
of Green functions with odd number of field operators.

\noindent \\ \textbf{CONCLUSION}

On the example of a real non-linear scalar field with the
interaction $\varphi^4$, a method of systematic description of
states of quantized fields with spontaneously broken symmetry is
proposed. The approach is based upon the use of the self-consistent
field model for the description of free particles. This model
accounts for the influence on the dynamics of the particles of the
average field formed by vacuum fluctuations. As distinct from the
Goldstone approach \cite{6} to whose results we come in the present
paper when neglecting the contribution of the vacuum average of the
square of field operator, the mass of free particles is determined
not only by the mass parameter entering into the initial Lagrangian,
but also by the interaction constant and the vacuum average from the
square of the field operators. As far as the integral which appears
in calculating the vacuum average diverges at large momenta, there
arises a need of introducing into the theory of a cutoff parameter
$\Lambda$. However, the non-linear equations for dimensionless
masses $m^2/\Lambda^2$ derived in the work already do not contain
explicitly this cutoff parameter, and the dimensionless particle
masses are determined in a unique way. There may exist not one, but
several solutions of the equations for masses at the same values of
parameters of the initial Lagrangian. These solutions correspond to
different states (phases) of the considered system. Among the
phases, there may exist both symmetrical phases and those with
broken symmetries. The most stable among them is the phase with the
lowest density of the vacuum energy. The density of the vacuum
energy of a scalar field is determined by both the positive
contribution of zero fluctuations and the negative contribution
conditioned by non-linear effects, so that the full density of the
vacuum energy of the Bose field can be either positive or negative.

An important advantage of the present approach is the fact that the
choice of the self-consistent field model for description of free
particles automatically brings us to the normal form of the
interaction Lagrangian. Owing to this fact, in all orders of the
perturbation theory all diagrams containing loops are excluded from
the consideration, and the structure of the perturbation theory is
substantially simplified. Also, the difficulties, conditioned by the
diagrams which contain the loops connected with one external line,
do not appear \cite{22}. The absence of loop diagrams in higher
orders of the perturbation theory means that their contribution is
already taken into account when constructing the zero approximation
Lagrangian that describes independent particles in the
self-consistent field model. As is known \cite{19}, the theories of
scalar field with interactions $\varphi^3$ and $\varphi^4$ are
renormalizable. The question of elimination of divergences in higher
orders of the perturbation theory, that is constructed on the basis
of the presented approach, is not discussed: this should be a
subject of a separate study. It should be noted, however, that as
far as the cutoff parameter is a substantial parameter of the theory
that eliminates the ultraviolet divergences, then, apparently, the
proposed approach can be also applied to non-renormalizable
theories.

In the present work it is shown that for calculation of averages
over the vacuum state of the exact Hamiltonian it is natural to use
the concept of quasi-averages that was earlier developed in the
framework of statistical mechanics \cite{16,21}. The equations for
exact one- and two-point Green functions are obtained.

The proposed method of description of the states with broken
symmetries is quite universal, because it lets consider on the basis
of the unified approach both Bose and Fermi fields, and also
describe the states with breaking of different symmetries, including
several symmetries at once \cite{23}.

\begin{center}
\refstepcounter{seksjon}
\renewcommand{\refname}{\normalsize References} 

\end{center}

\end{document}